\documentclass[12pt]{article}
\usepackage{graphicx}
\begin{document}
\centerline{\bf  Potts model with $q=4,6,$ and $8$ states on Voronoi-Delaunay random lattice$^{\star}$}

\bigskip
\centerline{F.W.S. Lima}

\bigskip
\noindent
Departamento de F\'{\i}sica,
Universidade Federal do Piau\'{\i}, 64049-550, Teresina - PI, Brazil \\
\medskip
  e-mail:  fwslima@gmail.com
\bigskip

$\star$ This paper is dedicated to Dietrich Stauffer \\

{\small Abstract: Through Monte Carlo simulations we study two-dimensional  Potts models with $q=4, \ 6$ and $8$ states
on Voronoi-Delaunay random lattice. In this study, we assume that the coupling factor $J$ varies with the distance $r$ between the first neighbors as $J(r)\propto e^{-a r}$, with $a \geq 0$ . The disordered system
is simulated  applying the singler-cluster Monte Carlo update algorithm and reweigting technique. In this model both
second-order  and first-order phase transition are present depending of $q$ values and $a$ parameter. The critical exponents ratio $\beta/\nu$, $\gamma/\nu$, and $1/\nu$ were calculated for case where the second-order phase transition are present. In the Potts model with $q=8$ we also studied the distribution of clusters sizes.}


\bigskip
{\bf Introduction}

This paper deals with the Potts model with $q=4, \ 6$ and $8$  states on two-dimensional Voronoi-Delaunay random lattices (VDRL). These lattices have a natural disorder in their coordination number. The randomness in the lattice of statistical spin models has been studied in order to access the effect of impurities and dilutions over their critical behavior. It was conjectured by Harris \cite{harris} that the sign of the critical exponent of the specific heat, $\alpha$, determines whether the system is affected or not by such a randomness. For positive values of $\alpha$ the impure system should have a critical behavior different from the one of the pure system. For negative values of $\alpha$, on the other hand, the critical behavior of the system should be same for both cases. 
In two-dimensional regular lattices, the ferromagnetic Potts model with $q$ states displays first order phase transitions for $q>4$ \cite{wu,Tsallis}. The pure ferromagnetic three-state Potts model has $\alpha=1/3$, hence, from the Harris criterium we expect to find a different behavior with a random interaction system. However, Picco \cite{Picco} and Lima et al. \cite{lima0} studied this model with different types of disorder and did not find significant differences from the pure case.
The $q$-state Potts model has been studied in scale-free networks by Igloi and Turban
\cite{igloi} and depending on the value of $q$ and of the degree-exponent  $\gamma$ first-
and second-order phase transitions are found. This model was also studied by Lima \cite{lima1} on {\it directed} Barab\'asi-Albert(BA) networks, where only one first-order phase transition has been obtained independent of $q$-values for connectivity $z=2$ and $z=7$ of the
{\it directed} BA network. Here, we studied the Potts model with $q=4$, 6, and 8 states. We also calculate the
critical exponents ratio $\beta/\nu$, $\gamma/\nu$ and $1/\nu$ for second-order phase transitions raised by disorder of the VDRL.

\bigskip

{\bf Model and  simulation}

The Voronoi construction or tessellation for a given set of points
in the plane is defined as follows \cite{christ}. Initially, for each point one
determines the polygonal cell consisting of the region of space
nearer to that point than to any other point. Then one considers that the
two cells are neighboring when they possess a boundary  in
common. From the Voronoi tessellation the dual
lattice can be obtained  by the following procedure:\\
$(a)$ when two cells are neighbors, a link is placed between the
two points located in the cells;\\
$(b)$ From the links one obtains the triangulation of space
that is called the Delaunay lattice;\\
$(c)$ The Delaunay lattice is dual to the Voronoi tessellation in
the sense that points corresponding to cells link to edges, and
triangles to the vertices of the Voronoi tessellation.\\

The Hamiltonian of an $q$-states ferromagnetic Potts model can be written as
\begin{equation}
-H=\sum_{<i,j>}J_{ij}\delta_{\sigma_{i}\sigma_{j}},
\end{equation}
where $\delta$ is the Kronecker delta function, the sum goes over all nearest-neighbors pairs of sites and the spin $\sigma$ can take the values $\sigma=1,..., q$. Here we assume that the coupling factor $J_{ij}$ depends on the relative distance $r_{ij}$ between sites $i$ and $j$ according to the following expression
\begin{equation}
J_{ij}=J_{0}e^{-ar_{ij}},
\end{equation}
where $J_{0}$ is a constant, set equal to unity for simplicity, and $a\geq 0$ is a model parameter.

The simulations have been performed  applying the single-cluster update algorithm \cite{sc} on different lattice sizes  comprising a number
$N=250$, $1000$, $2000$, $4000$, and $8000$ of sites. For simplicity, 
the length of the system is defined
here in terms of the size of a regular lattice $L=N^{1/2}$.
For each system size quenched averages over the connectivity disorder are 
approximated by averaging over $R=20$ 
independent realizations. For each
simulation we have started with a uniform configuration of spins
(the results are however independent of the initial configuration). 
We ran $2.52\times10^{6}$ Monte Carlo steps (MCS) per spin with $1.2\times10^{5}$ configurations 
discarded for thermalization using the "perfect" random-number generator \cite{nu}.

In studying the critical behavior of the model using  the single-cluster algorithm we define the variable $e=E/N$,
where $E$ is the energy of system, and the magnetisation of system  $M=(q.\max[n_{i}]-N)/(q-1)$ in a time series file, where $n_{i}\leq N$ denote the number of spins with "orientation" $i=1,...,q$. From the fluctuations of $e$
measurements we can compute: the average of $e$, the specific heat $C$ and the
fourth-order cumulant of $e$,
\begin{equation}
 u(K)=[<E>]_{av}/N,
\end{equation}
\begin{equation}
 C(K)=K^{2}N[<e^{2}>-<e>^{2}]_{av},
\end{equation}
\begin{equation}
 B(K)=[1-\frac{<e^{4}>}{3<e^{2}>^{2}}]_{av},
\end{equation}
where $K=1/k_BT$, $T$ is the temperature, and $k_B$ is the Boltzmann constant.
Similarly, we can derive from the magnetization measurements
the average magnetization ($m=M/N$), the susceptibility, and the magnetic
cumulants,
\begin{equation}
 m(K)=[<|m|>]_{av},
\end{equation}
\begin{equation}
 \chi(K)=KN[<m^{2}>-<|m|>^{2}]_{av},
\end{equation}
\begin{equation}
 U_{4}(K)=[1-\frac{<m^{4}>}{3<m^{2}>^{2}}]_{av}.
\end{equation}
where $<...>$ stands for a thermodynamic average and $[...]_{av}$ square brackets
for an average over the 20 realizations.

In order to verify the order of the transition for this model, we apply finite-size scaling
(FSS) \cite{fss}. Initially we search for the minima of the fourth-order parameter of Eq.
(4). This quantity gives a qualitative as well as a quantitative description of the order
of the transition \cite{mdk}. It is known \cite{janke} that this parameter takes a minima
value $B_{\min}$ at effective transition temperature $T_{c}(N)$. One can show \cite{kb}
that for a second-order transition $\lim_{N\to \infty}$ $(2/3-B_{\min})=0$, even at
$T_{c}$, while at a first-order transition the same limit measuring the same quantity is
small and $(2/3-B_{\min})\neq0$.

A more quantitative analysis can be carried out through the FSS of the $C$ fluctuation
$C_{\max}$, the susceptibility maxima $\chi_{\max}$ and the minima of the Binder parameter
$B_{\min}$. 

If the hypothesis of a first-order phase transition is correct, we should
then expect, for large systems sizes, an asymptotic FSS behavior of the form
\cite{wj},
\begin{equation}
C_{\max}=a_{C} + b_{C}N +...
\end{equation}
\begin{equation}
\chi_{\max}=a_{\chi} + b_{\chi}N +...
\end{equation}
\begin{equation}
B_{\min}=a_{B} + b_{B}/N +...
\end{equation}

Therefore, if the hypothesis of a second-order phase transition is correct, we should
then expect, for large systems sizes, an asymptotic FSS behavior of the form
\begin{equation}
 C=C_{reg}+L^{\alpha/\nu}f_{C}(x)[1+...],
\end{equation}
\begin{equation}
 m=L^{-\beta/\nu}f_{m}(x)[1+...],
\end{equation}
\begin{equation}
 \chi=L^{\gamma/\nu}f_{\chi}(x)[1+...],
\end{equation}
\begin{equation}
\frac{dU_{4}}{dT}=L^{1/\nu}f_{U}(x)[1+...],
\end{equation}
\begin{equation} 
|\frac{d\,\ln\,m^p}{d\,K}|_{\max}\,=\,|\frac{dm^p/dK}{m^p} 
|_{\max} \propto \,L^{p/\nu}[1+...] \label{p2}, 
\end{equation}
where $C_{reg}$ is a regular background term,
$\nu$, $\alpha$, $\beta$, and $\gamma$ are the usual critical
exponents, and $f_{i}(x)$ are FSS functions with
\begin{equation}
 x=(K-K_{c})L^{1/\nu}
\end{equation}
being the scaling variable, and the brackets $[1+...]$ indicate
corretions-to-scaling terms. Therefore, from the size dependence of $M$ and $\chi$
we obtained the exponents $\beta/\nu$ and $\gamma/\nu$, respectively. The exponent
$1/\nu$ is obtained from the relations (15) and (16).
The maxima value of susceptibility also scales as $L^{\gamma/\nu}$. 
\bigskip
\newpage
\begin{figure}[ht]
\begin{center}
\includegraphics [angle=0,scale=0.4]{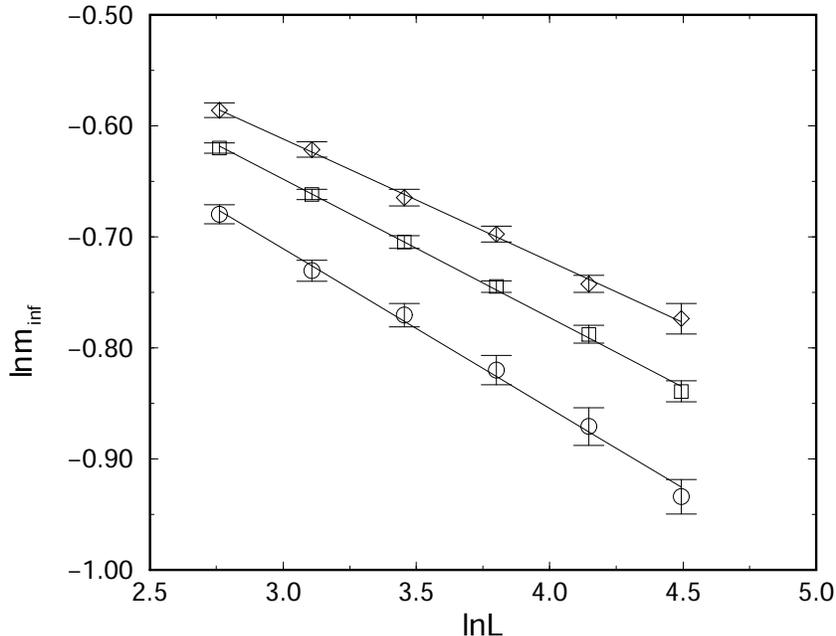}
\end{center}
\caption{Logarithmic plot of the magnetisation at the inflection point versus the size system $L$ for $a=0.0$ (circles),
$0.5$ (squares) and $1.0$ (diamonds), $q=4$.}
\end{figure}
\begin{figure}[ht]
\begin{center}
\includegraphics [angle=0,scale=0.4]{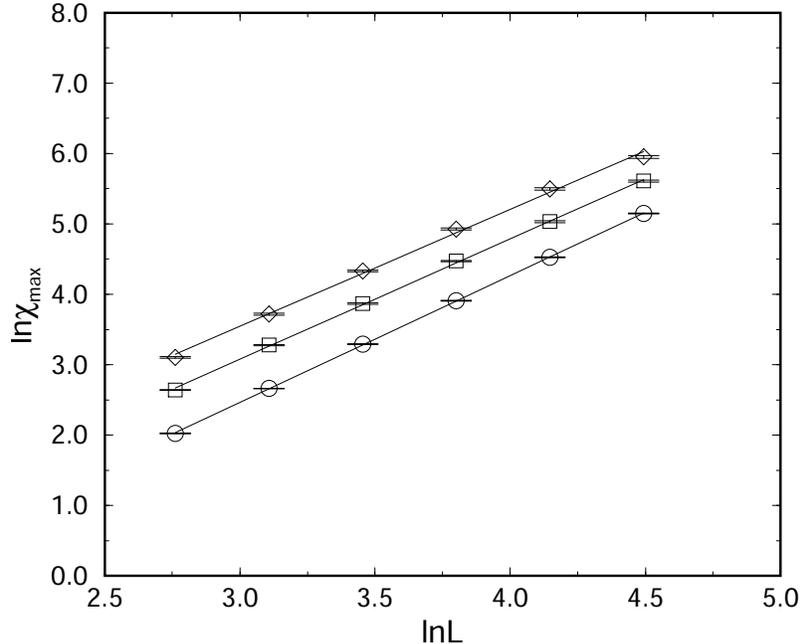}
\end{center}
\caption{Logarithmic plots of the maximum suscepbility versus the
size  system $L$ for $a=0$ (circles),
$0.5$ (squares) and $1.0$ (diamonds), $q=4$.}
\end{figure}
\begin{figure}[ht]
\begin{center}
\includegraphics [angle=0,scale=0.4]{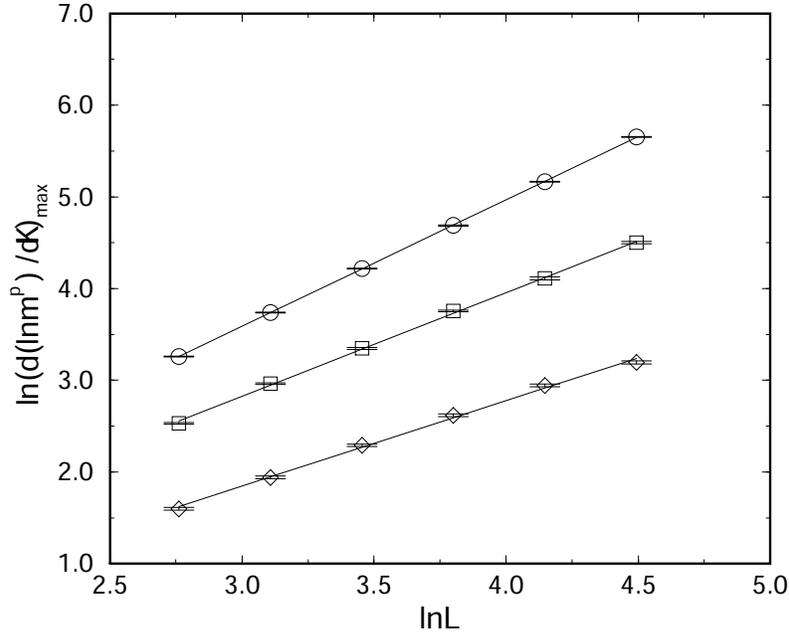}
\end{center}
\caption{Plots of scale of finite size of the maximums of the Logarithmic derivatives of $|M|^{p}$ versus $\ln L$ for $p=2$ and $a=0$ (circles),
$0.5$ (squares) and $1.0$ (diamonds), $q=4$.}
\end{figure}
\begin{table}[h]
\begin{center}
\begin{tabular}{|c|c|c|c|c|}
\hline
$ $ & $\nu$ & $\alpha/\nu$& $\beta/\nu$ & $\gamma/\nu$ \\
\hline

$ (i)$& $2/3$ & $1$ & $1/8$ & $7/4$  \\
$     $ & $(0.6666)$ & $ (1)$ & $ (0.125) $ & $(1.75) $   \\
\hline
$ (ii)\hspace{0.2cm} a=0  $    & $0.725(2)$ & $0.713(4)$& $0.143(9)$ &
$1.799(6)$  \\ 
$           $    & $0.727(6)$ & $$& $$ & $$ \\ 
$  \hspace{0.95cm}    a=0.5$    & $0.88(1)$& $0.43(2)$& $0.12(2)$ &
$1.70(2)$  \\
 $           $    & $0.88(2)$ & $$& $$ & $$ \\
$ \hspace{0.95cm}     a=1.0$    & $1.07(3)$ & $0.22(1)$& $0.10(2)$ &
$1.66(4)$  \\ 
$           $    & $1.07(2)$ & $$& $$ & $$ \\

\hline
\end{tabular}
\end{center}
\caption{Ferromagnetic of Potts with $4$ states in two-dimensions
: (i) analytical results, (ii) our results for $a=0.0$, 0.5 and 1.0. Where have two values
for $\nu$ exponent calculated using the eq. (16) for $p=1$ and $2$.}
\label{table1}
\end{table} 

\begin{figure}[ht]
\begin{center}
\includegraphics [angle=0,scale=0.4]{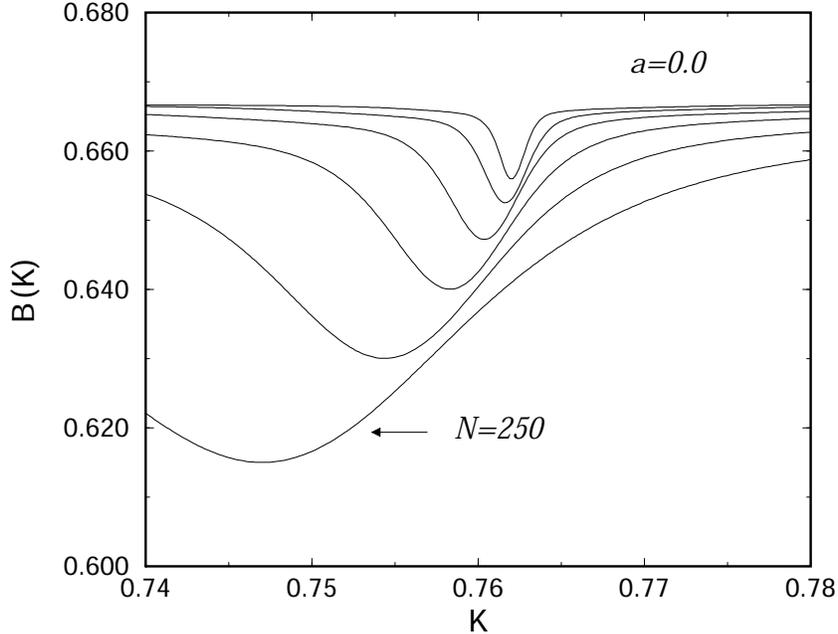}
\end{center}
\caption{Plots of the Binder parameter $B(K)$ versus $K$ for $a=0$
and several lattices sizes ($N=250$, 500, 1000, 2000, 4000,
 and 8000) sites, for $q=6$.}
\end{figure}

\begin{figure}[ht]
\begin{center}
\includegraphics [angle=0,scale=0.4]{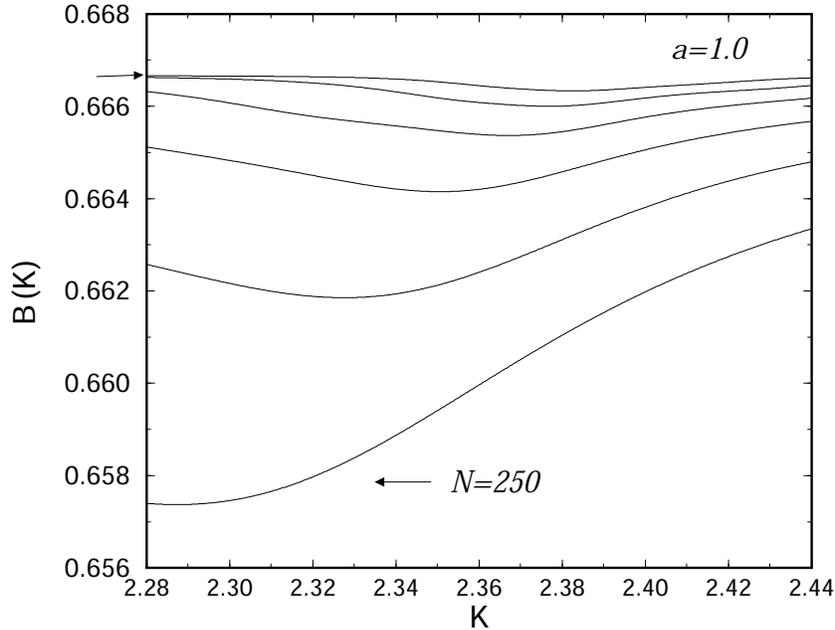}
\end{center}
\caption{The same as in Figure 4, but for $a=1$. The arrow indicates
the position of $B(K)=2/3$.}
\end{figure}

\begin{figure}[ht]
\begin{center}
\includegraphics [angle=0,scale=0.4]{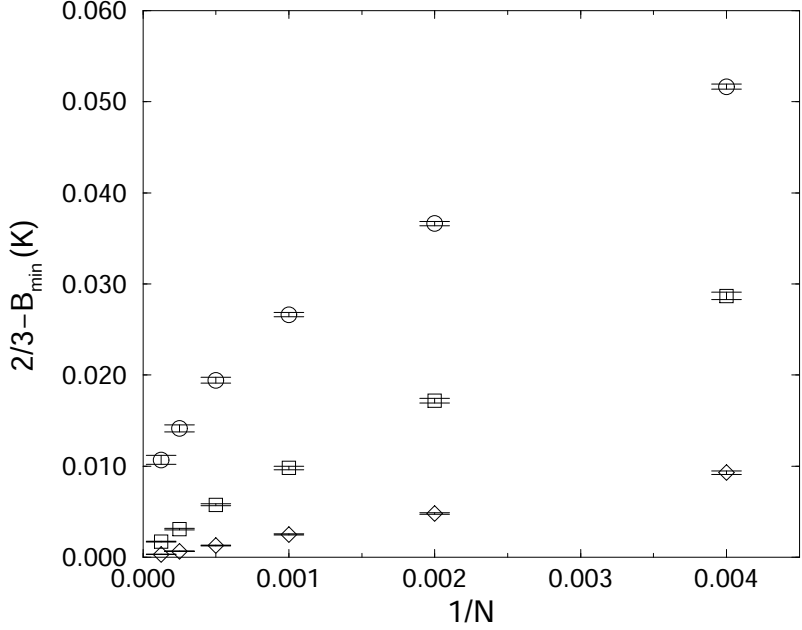}
\end{center}
\caption{Plots of $2/3-B_{\min}(K)$ versus $1/N$ for $a=0$ (circles),
$0.5$ (squares) and $1.0$ (diamonds), $q=6$.}
\end{figure}

\begin{figure}[ht]
\begin{center}
\includegraphics [angle=0,scale=0.4]{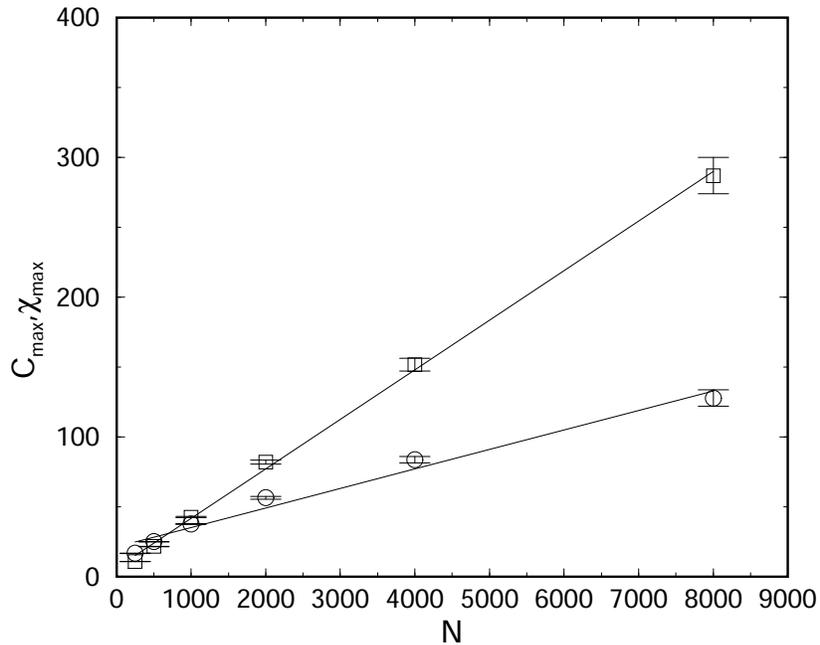}
\end{center}
\caption{Plots of the specific heat $C_{\max}$(circles) and susceptbility
$\chi_{\max}$(squares) versus $N$ for $a=0$, $q=6$.}
\end{figure}
\begin{table}[h]
\begin{center}
\begin{tabular}{|c|c|c|c|c|}
\hline
$ $ & $\nu$ & $\alpha/\nu$& $\beta/\nu$ & $\gamma/\nu$ \\
\hline
$ (i)$ & $1$ & $0$& $0.125$ & $1.75$  \\
\hline
$ (ii)$ & $1.03(3)$ & $0.182(5)$ & $0.120(9)$ & $1.750(6)$  \\
$     $ & $1.04(3)$ & $ $ & $     $ & $    $   \\
\hline 
$(iii)\hspace{0.2cm} a=0.5$    & $0.93(5)$& $0.51(5)$&
$0.122(4)$ & $1.53(5)$  \\ 
$           $    & $0.91(5)$ & $$& $$ & $$
\\
$      \hspace{0.95cm} a=1.0$    & $1.11(4)$ & $0.22(2)$&
$0.14(1)$ & $1.56(5)$  \\ 
$           $    & $1.10(4)$ & $$& $$ &
$$ \\

\hline
\end{tabular}
\end{center}
\caption{ (i) Analytical results for ferromagnetic Ising model $2D$, (ii) Simulation results of Janke
{\it et. al.} \cite{Jan94} for ferromagnetic Ising model $2D$ on Voronoi-Delaunay random lattice.  (iii) Our 
results for ferromagnetic $2D$ Potts model with $q=6$ for $a=0.5$ and 1.0. Again we have two values for exponent $\nu$, see eq. (16) and table 1.} 
\end{table}

\begin{table}[h] 
\begin{center} 
\begin{tabular}{|c||c|c|c|c|c|} 
\hline 
$a$ & $\nu_{1}$ & $\nu_{2}$ & $\alpha/\nu$& $\beta/\nu$ & $\gamma/\nu$\\ \hline 
$0.5$ & $1.22(5)$ & $1.17(3)$ & $0.32(7)$& $0.131(7)$ & 
$1.20(8)$ \\ \hline 
$1.0$ & $1.26(3)$ & $1.26(8)$ & $0.12(3)$& $0.126(8)$ & 
$1.45(6)$ \\ \hline 
\end {tabular} 
\end{center} 
\caption{Critical exponents for ferromagnetic Potts with $8$ states in $2D$ on VRDL. Our  
results for $a=0.5$ and $1.0$. The $\nu_{1}$ and $\nu_{2}$ exponents are respectively for
$p=1$ and $2$, see eq. (16). }  
\end{table} 

\begin{figure}[ht]
\begin{center}
\includegraphics [angle=0,scale=0.4]{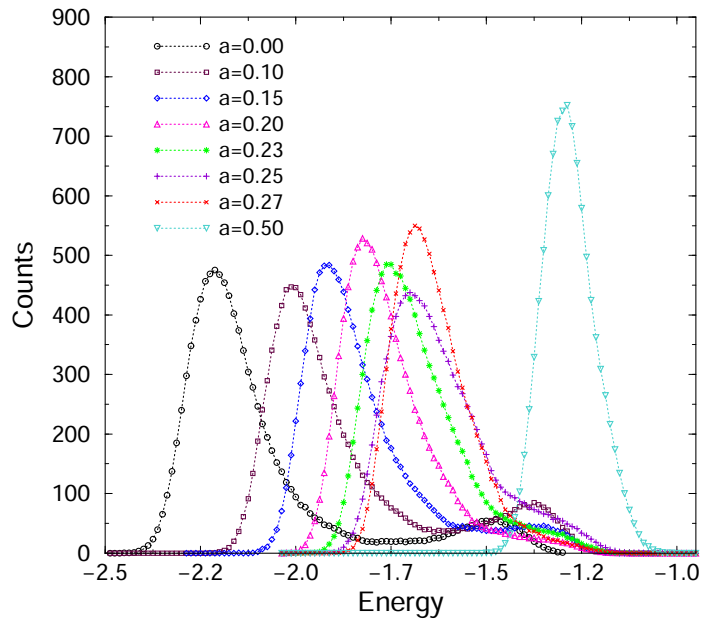}
\end{center}
\caption{Energy histogram for $a=0.0$ (left) to $0.5$ (right) and $N=4000$ 
sites, $q=8$.}
\end{figure}

\begin{figure}[h]
\begin{center}
\includegraphics [angle=0,scale=0.4]{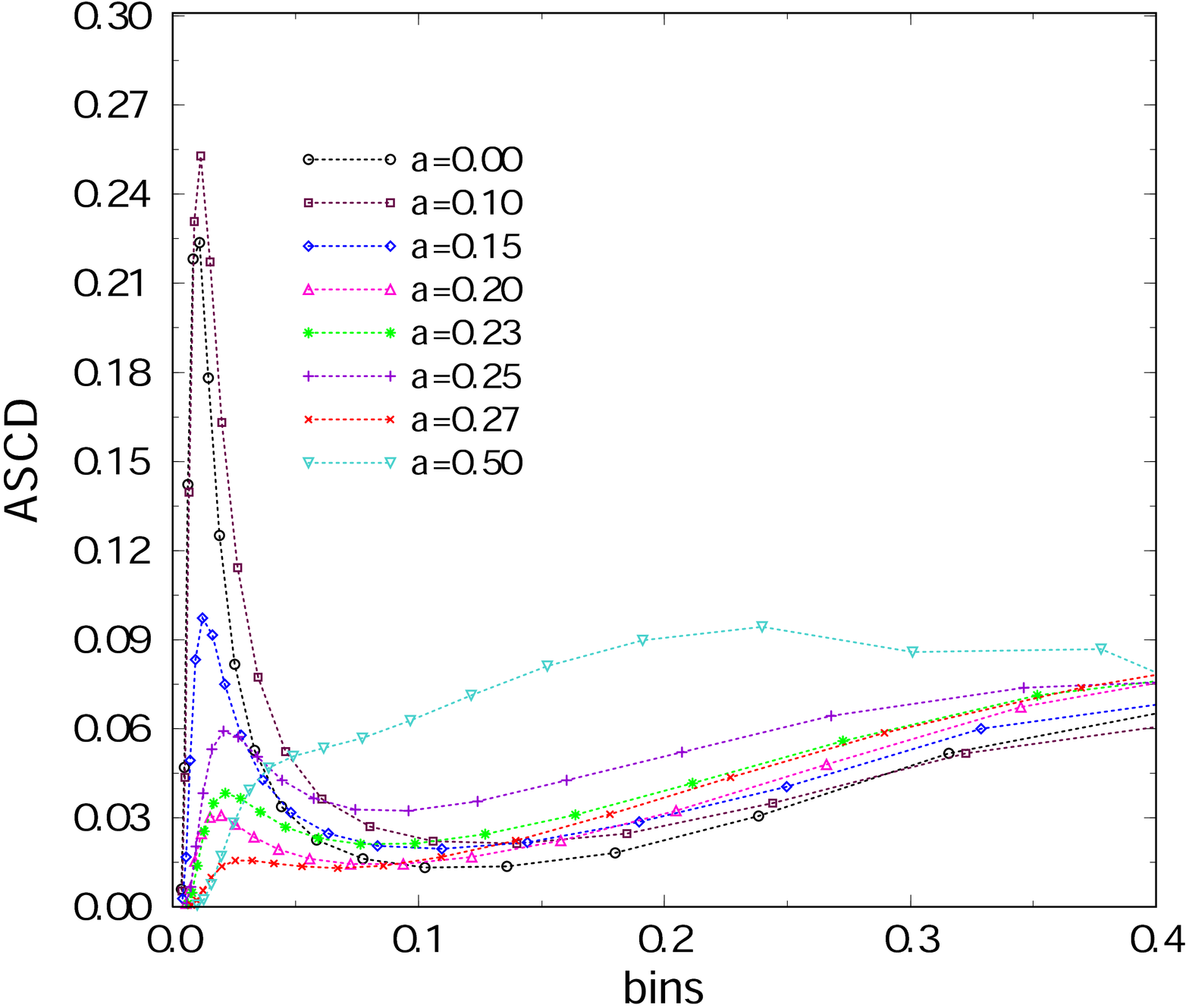}
\end{center}
\caption{Configuration averaged cluster size distribution (ASCD) for $a=0.0$
to $0.5$ and $N=4000$ sites, $q=8$.}
\end{figure}
{\bf  Cluster distributions in two-dimensional $8$-states
Potts model on VDRL}

To distinguish the order of the phase transition 
is one of various problems of Monte Carlo simulations
for spin systems. The greatest difficulty
arises when the correlation length
is finite and larger than the size of the system.. 
In such situations, common tools
identification of the order of transition as
for example, by examining the minimum
free energy \cite{lee90, lee91} by considering
the probability distribution of energy \cite{Bin86} 
may not be in this case, a good indicator.
Even in the presence of meta-stable states,
the size of the system can prevent the observation of the double peak
characteristic of the first-order phase transition,
in the distribution of energy. 
The basic reason for such behavior occurs because
the fact that energy is a local quantity.
Quantities of a global nature are expected to be
more sensitive to the correlation length
and so the effects of meta-stable states
become more evident. 
 
If a system undergoes a second-order phase transition, the correlation length
is infinitely large at the critical point of the system
which results in the formation of infinitely
large clusters, such that  close to the point of
a critical second-order phase transition there is co-existence 
of clusters of all sizes.  
 
In systems that exhibit first-order phase transitions
 the most common feature
is the coexistence of ordered and disordered states
in the region of phase transition.
While structures dominated by large clusters are ordered phases,
small clusters are representative of the 
disordered phase. The existence of both small
and large clusters at a first-order phase transition 
 gives rise to the behavior of
double peaks in the energy distribution. 
 
Previous works
\cite{aydin96,aydin97} 
on the Potts model in two dimensions
observed that global operators related to
clusters size are more sensitive
structural changes in a phase transition 
that operators of sites related to energy
and the order parameter of the system. Particularly
the cluster distribution may give a better
indication of the order of phase transition
for small network sizes than the energy distribution. 
The average clusters size $S$ may
be defined \cite{fatih} as 
\begin{equation}
 S=\frac{1}{N_{c}}\langle \sum_{i=1}^{N_{c}}C_{i}\rangle,
\end{equation}
where $N_{C}$ is the number of clusters 
and $C_{i}$ the spin number in the cluster 
$i$ normalized by the total spins number. One possible
cluster related measure is the cluster size distribution (CSD),
which is evaluated in the same way as the standard
energy probability distribution in the form of histogram. Another
relevant quantity is the configuration average cluster size
distribution (ASCD). Instead of counting all clusters, ASCD is
obtained by considering average cluster sizes for each configuration.

\bigskip

\newpage

{\bf Results and Discussion}

We study the critical behavior of the Potts model on VDRL for                                                    three values of $a$ ($a = 0.0$, $0.5$ and $1.0$) and three values $q$ ($q=4$, $6$, and
$8$). For each value of $a$ and $q$, we apply the finite size scaling technique \cite{fss} together with the single-histogram algorithm. We perform the same procedure for systems with different number of sites $N = 250$, 500, 1000, 2000, 4000, and 8000. The critical temperature for infinite size system is estimated by using the fourth-order magnetization (Binder) cumulant.

In Figs. 1 to 3 we display the scalings for natural logarithm for the dependence of the magnetization $|m|$  on inflection point at $K=K_{c}(L)$,  maximum amplitude, $\chi_{\max}(L)$, and logarithmic derivatives of $|M^{p}|$($p=2$) versus $\ln L$ for $a = 0.0$ (circles), $0.5$ (squares) and $1.0$ (diamonds), respectively, for $q=4$. Using the fourth-order magnetisation Binder cumulant we find for $q=4$ the critical values $K_{c} = 0.6691(3)$, $1.2092(5)$ and $2.1353(4)$ and $U^{*}= 0.6031(2)$, $0.6199(4)$ and $0.6272(3)$, corresponding to $a = 0.0$, $0.5$ and $1.0$, respectively. For $q=4$ the specific-heat exponent $\alpha=2/3$ is good candidate to check whether a change of critical behavior can be induced at all by VDRL. Here, our results, see table 1, presents no {\it reliable} indication for change and in favor of the  Harris criterium,  error bars are only statistical, and much larger systems might give different exponents, also, the scaling laws
$2-\alpha = \gamma + 2\beta$ $=$$d\nu$ are partially violated, making our estimates less reliable.

For the Potts model with $6$-states and $a=0$ ($K_{c} = 0.7633(6)$) our simulations indicate that the model display a first-order phase transition, in perfect agreement with the results reported by Janke et al. \cite{wj}. However at high value of $a$, $a=0.5$ and $1.0$, we observe a typical second-order phase transition where the critical values $K_{c} = 1.3671(5)$ and $2.3972(6)$  and $U^{*}= 0.6236(6)$ and $0.6324(5)$, corresponding to $a = 0.5$ and $1.0$, respectively, was obtained using the eq. 8. 

In Figs. 4 and 5 we display the plot $B(K)$ versus $K$ for $a=0.0$ and $1.0$
and different size lattices ($N=250$ to $8000$ sites). We can see that, in the limit
of large lattice size the Binder parameter goes to $2/3$ (in Fig. 5), providing a qualitative confirmation for the presence of the continuos transition in the system.
For $a=0$ (see Fig. 4), however, the Binder parameter goes to a value which is different from $2/3$. This is a sufficient condition to characterize a first-order  phase transition. The order of the transitions can be confirmed by plotting the values of 
$2/3-B_{\min}(K)$ versus $1/N$ for different values of $a$. While for $a=0.5$ and $1.0$
the curve goes to zero as we increase the system size, for $a=0$ the quantity
$2/3-B_{\min}(K)$ approaches a nonvanishing value in the limit of small $1/N$
(see Fig. 6). At this point, we can assume that change in $a$ from $0.0$ to $0.5$
should be followed by a crossover at a value $a=a_{c}$ from a first- to second-order phase transitions. A  quantitative analysis can be made through of the relations (9) and (10), see Fig. 7. The exponents ratio $\beta/\nu$, $\gamma/\nu$ and $1/\nu$ are obtained from the slopes of the straight lines for $q=6$ and $a=0.5$ and $1.0$ (not shown here) the same way was obtained for $q=4$, see table 2. For Potts model with $q=8$ states Lima et al. \cite{wel} have  made identical numerical analysis made here for $q=6$, see table 3. Here we simulate, using
cluster algorithms, the Potts model
studied in previous sections, observing energy histograms and the ASCD
in order to obtain better
information on the phase transition of this model
and speculate on the value of $ a_ {c} $ mentioned
previously. 
The cluster algorithm used 
here is the same  Wolff algorithm used previously,
 except that before
to calculate the observables, we studied
clusters until the total number of
sites in all clusters visited
is equal to or greater than the total number
sites of the network. I.e., while in the conventional Wolff algorithm only one
cluster is formed in one Monte Carlo interaction, here we can have more than
one  cluster formed. 
In our simulation we
study the ferromagnetic Potts model
with eight states at $a=0.00,$
0.10, 0.15, 0.20, 0.23, 0.25, 0.27 and 0.50 for $N=4000$ 
sites. After thermalization of 
300,000  Monte Carlo steps about $10^{6}$ steps were
simulated in their respective transition points.
After every two steps, we computed the energy and the
average cluster size. 
 
We examined the phase structure of the 
ferromagnetic  Potts model with eight states
with respect to  quenched randomness in
a range of values of $ a $ from $ a = 0.00$ to  $ 0.50$. For a fixed value of $ a $,
 $40 $ replicas were
generated with different bounds distributions. 
Energy histograms and ASCD were obtained from
averages of those obtained in the above
$ 40 $ replicas generated. In Fig. 8, 
we show the energy histogram over a network
of $N=4000$ sites at various values of $a$. 
Starting with $a=0.00$, which displays a first-order
phase transition for a ferromagnetic
Potts model with eight states studied here, we see 
a double peak structure in the energy histogram 
for values of $a$ below $a=0.20$. For values 
of $a$ above $a=0.25$
($x$ represents $a=0.27$
and the triangles correspond to $a=0.50$ in the Fig. 9)
we observed a
single Gaussian peak in the energy histogram. To
obtain a better signal with respect to the type
of the transition, we examined the ASCD. As shown in Fig. 9,
the peaks in the region of small clusters 
indicate the presence of first-order phase transitions. 
The order of the transition
changes from first to second order in the range of 
$0.20 \leq a \leq 0.27$, where the exact value of  
$a_c$, crossover, was not obtained here, we
even in this range observed the presence of meta-stable 
states for the size of the network studied. 
What remains is to verify whether the threshold value
of crossover for this type of randomness is
specific or not for the size of the network studied.

\bigskip 
\bigskip
\newpage
\newpage
{\bf Conclusion}

In conclusion, we have presented simulations for Potts model with  $q=4$, $6$ and $8$ states on VDRL. The disordered system is simulated  applying the singler-cluster Monte Carlo update algorithm and reweigting technique that give results with precision hight.
The  Potts model with $q=4$ does display a second-order phase
transition on VDRL for parameter $a=0$, $0.5$, and $1.0$. For the Potts model on regular
lattice the specific-heat exponent $\alpha=2/3$ is a good candidate for a change of critical exponents on VDRL. Here, our results, summarized in table 1, presents no {\it reliable} indication for change, because the  error bars are only statistical. Nevertheless, the present exponent estimates give a change, in particular for $\alpha/\nu$, and then are compatible with the Harris criterium for $N=250, 500$,
$1000$, $2000$, $4000$, and $8000$ sites used here and agree with the assumption made by Janke et al. \cite{jankeB} that the Harris criterium is not violated for $q=4$.
For $q=6$ and $8$ with $a=0$ this model presents a first-order phase that is  agreement with results for regular lattices, i.e., connectivity disorder only not is enough to change the order of phase transition that agree with results of Janke and Villanova \cite{jv}.
For $a>0$ it presents second-order phase
transition. Thus, the above results, summarized in tables  2 and 3 show
that the Potts model case studied here on VDRL
is similar to the critical behavior of the two-dimensional ($2D$) eight-state random-bond Potts model, S. Chen et al. \cite{schen}. They obtained the critical exponents for
two sets of bond strengts, from which they concluded  that the transition is the second order with critical exponents for both sets falling into same universality class, that 
of a $2D$ Ising model, i.e., $\alpha/\nu=0$, $\beta=1/8=0.125$,
$\gamma=7/4=1.75$ and
$\nu=1$. Here again the exponents ratio $\alpha/\nu$ for $q=6$ and $8$ 
($a>0$) are large compared to the zero for square lattice Ising model and as
our disorder system is not the result
of the introduction of impurities in the pure system
nothing can be stated about the universality class of a $2D$ Ising model. However, the results of de Oliveira et al. \cite{dickman} studying the Contact Process on a Voronoi triangulation are in disagreement with the Harris criterium; these authors suggest then that the Harris criterium must be reformulated. The exponent $\gamma/\nu$ for $q=6$ and $8$ states and in the cases where $a>0$ do not change with the disorder of the Voronoi-Delaunay random lattices, but nothing can be said about the Harris criterium. Here, for Potts model
with eight states we also show that using quantities depending on cluster size distribution may
give complementary indications for identifying the order the phase transition
for small random lattice as VRDL, which may not be derivable from energy and
order parameter distribution.

The author thank  D. Stauffer for many suggestions and fruitful discussions during the
development this work and also for the revision of this paper. We also acknowledge the
Brazilian agency FAPEPI (Teresina-Piau\'{\i}-Brasil) for  its financial support. This
work also was supported the system SGI Altix 1350 the computational park
CENAPAD.UNICAMP-USP, SP-BRAZIL.

\end{document}